\documentclass[aps,prd,reprint,superscriptaddress,nofootinbib]{revtex4-2}
\usepackage{aas_macros}
\usepackage{graphicx}
\usepackage{hyperref}
\usepackage{amsmath}

\usepackage[dvipsnames,table]{xcolor}
\usepackage{hyperref}
\hypersetup{urlcolor=RedViolet,citecolor=ForestGreen,linkcolor=RoyalBlue,colorlinks=true}
\newcommand{\ie}{\emph{i.e.}}

\begin{document}

%Title of paper
\title{Constrained simulations of the local Universe with Modified Gravity}

\author{Krishna Naidoo}
\email[]{knaidoo@cft.edu.pl}
\affiliation{Center for Theoretical Physics, Polish Academy of Sciences, al. Lotnik\'{o}w 32/46 Warsaw, Poland}
\affiliation{Department of Physics \& Astronomy, University College London, Gower Street, London, WC1E 6BT, UK}
\author{Wojciech A. Hellwing}
\email[]{hellwing@cft.edu.pl}
\affiliation{Center for Theoretical Physics, Polish Academy of Sciences, al. Lotnik\'{o}w 32/46 Warsaw, Poland}
\author{Maciej Bilicki}
\affiliation{Center for Theoretical Physics, Polish Academy of Sciences, al. Lotnik\'{o}w 32/46 Warsaw, Poland}
\author{Noam Libeskind}
\affiliation{Leibniz-Institut f\"{u}r Astrophysik Potsdam, An der Sternwarte 16, D-14482 Potsdam, Germany}
\affiliation{University of Lyon, UCB Lyon 1, CNRS/IN2P3, IUF, IP2I Lyon, France}
\author{Simon Pfeifer}
\affiliation{Leibniz-Institut f\"{u}r Astrophysik Potsdam, An der Sternwarte 16, D-14482 Potsdam, Germany}
\author{Yehuda Hoffman}
\affiliation{Racah Institute of Physics, Hebrew University, Jerusalem 91904, Israel}

\date{\today}

\begin{abstract}
We present a methodology for constructing modified gravity (MG) constrained simulations of the local Universe using positions and peculiar velocities from the CosmicFlows data set. Our analysis focuses on the following MG models: the normal branch of the Dvali-Gabadadze-Porrati (nDGP) model and Hu-Sawicki $f(R)$ model. We develop a model independent methodology for constructing constrained simulations with any given power spectra and numerically calculated linear growth functions. Initial conditions (ICs) for a set of constrained simulations are constructed for the standard cosmological model $\Lambda$CDM and the MG models. Differences between the model's reconstructed Wiener filtered density and the resultant simulation density are presented showing the importance for the generation of MG constrained ICs to study the subtle effects of MG in the local Universe. These are the first MG constrained simulations ever produced. The current work paves the way to improved approximate methods for models with scale-dependent growth functions, such as $f(R)$, and for high-resolution hydrodynamical MG zoom-in simulations of the local Universe.
\end{abstract}

% insert suggested keywords - APS authors don't need to do this
%\keywords{}

\maketitle

\section{Introduction\label{intro}}

The standard model of cosmology, $\Lambda$ + cold dark matter ($\Lambda$CDM) stands on firm pillars formed by a multitude of observational tests. From the early epochs of primordial nucleosynthesis \citep{Walker1991}, and the statistical properties of the cosmic microwave background radiation and its angular fluctuations \citep{Planck2020}, to the late time formation and evolution of large scale structure as gloriously manifested by the amazing spatial patterns observed from vast and deep modern-age galaxy surveys (such as the Baryon Oscillation Spectroscopic Survey \citep{BOSS2015}). The observations concerning all of these epochs and phenomena can be explained remarkably well by a simple 6-parameter $\Lambda$CDM model. These undeniable empirical successes that form the foundations of $\Lambda$CDM come however mostly either from the large scales or early epochs when the physics of the involved phenomena is generally in the linear or mildly-nonlinear regimes.

In the past few decades the precision, quality and volume of the data concerning the so-called local Universe has grown and improved many-folds. Thus opening the window on non-linear scales, where the environment defined by the large-scale matter/galaxy distribution is entangled with local non-linear processes driving the evolution of galaxies, and their motions and clustering. Dedicated observational campaigns have brought us a detailed and interesting picture of our close cosmic neighbourhood -- within a region of 100-200 $h^{-1}{\rm Mpc}$ from the Local Group. With new observations of hundreds and thousands of local dwarf galaxies, more precise measurements of the Virgo, Coma and other local galaxy clusters to the stunning Cosmic Flows data containing distances and velocities of tens of thousands of nearby galaxies. All these new local Universe data have a great potential for providing new stringent tests for cosmology and new insights into non-linear phenomena and physics of galaxy formation and evolution. They also provide new intricate details concerning the local Universe that require robust modelling and understanding of structure formation physics in the non-linear regime.

Classically, the high-resolution state-of-the-art cosmological simulations provide a powerful tool for modeling and understanding 
the physics of the non-linear structure formation regime \citep{Davis1985}. However, since classical simulations use random phases of the initial conditions, the structures reflect just one particular random realisation from a vast ocean of potential configurations. Aiming at simulating the very local universe requires running very high resolution simulations, to capture the intricate physics in the deeply non-linear regime. At the same time one needs to run large box simulations, in order to sample Local Group-like environments out of cosmic variance. Large box high resolution galaxy formation simulations are prohibitively expensive.

Here, the constrained simulations \citep{CR1991} of the local Universe which reproduce structures and phase information from observations, allows us to circumvent the aforementioned limitations of random-phase simulations. These new class of carefully designed and engineered simulations nowadays play a critical role in distinguishing the subtle features from model extensions to $\Lambda$CDM. In constrained simulations the variance of the structures and resultant statistics is greatly reduced -- a property which is especially important in the context of the local Universe where observations of galaxies are made to higher accuracy (than those made on distant galaxies typically measured from galaxy redshift surveys) and there is greater sensitivity to small scale physics (such as the properties of dwarf galaxies) that is not available to larger surveys. Simulations mimicking structures from real galaxy observations have been produced from a variety of methodologies; constrained simulations \citep{CR1991} using peculiar velocities constraints from the local Universe \citep{CLUES2010, Sorce2016, HESTIA2020} (the focus of this paper), Bayesian reconstruction of the density field \citep{BORG2013,ELUCID2016} and methods using both density and peculiar velocity constraints \citep{Lilow2021}.

In this paper we focus on constructing constrained simulations of the local Universe using peculiar velocities \citep{Zaroubi1999}. Unlike estimations of the density field from galaxy positions -- biased tracers of the density field -- peculiar velocities in the linear regime are related directly to the density field. This allows us to estimate the density field with limited dependence on the galaxy tracer bias, while the bias from non-linear and stochastic galaxy motions are limited by considering only large scale linear regime reconstructions of the density field. Furthermore, peculiar velocities come with the following limitations: (1) measurements are limited to only the radial component of the velocities, (2) they can only be measured at low redshift (depending on accurate distance measurements and assumptions on the background expansion and Hubble constant) (4) non-linear velocities need to be removed and (5) Malmquist bias need to be considered. See \cite{Strauss1995} for a more detailed review. As a result constrained simulations from peculiar velocities have been limited to the local environment, in particular using data from CosmicFlows \citep[see][]{CF4}. Several studies, such as the Constrained Local UniversE Simulations (CLUES) \citep{CLUES2010, Sorce2016} and the High-resolution Environmental Simulations of The Immediate Area (HESTIA) \citep{HESTIA2020}, have used peculiar velocities from CosmicFlows to produce constrained simulations of the local Universe. However, these simulations have been limited to $\Lambda$CDM and the model dependence of the reconstruction and the reproduced structures has yet to be explored.

Modified gravity (MG) are a class of models that extend Einstein's theory of general relativity. They are motivated by a desire to provide a better theoretical explanation of the cosmological constant, a constituent of nature that has proven to be a significant challenge for theoretical fundamental physics. They achieve this by introducing an effective fifth force, that replicates the effects of the cosmological constant but that is screened on relatively small cosmological scales. The range of this screening mechanism and the high fidelity observations of the local Universe, make it a perfect environment for extracting the subtle effects of MG. In this paper we focus on the following MG models: Hu-Sawicki $f(R)$ with chameleon screening \citep{FofR} and the normal branch of the Dvali-Gabadadze-Porrati (nDGP) with Vainshtein screening \citep{nDGP}. Constrained simulations have never been produced for models of MG and those produced for the local Universe would enable detailed comparison studies looking to smaller scales than is typically accessible to large cosmological galaxy surveys.

In this paper we extend the methodology used by CLUES for creating constrained local Universe simulations to two models of MG. We particularly focus on making these methods numerically driven, enabling the extension to other models in future studies. The methods used by CLUES follows the procedures in \citep{Doumler1, Doumler2, Doumler3} which are implemented in \texttt{ICeCoRe} (Initial Conditions \& Constrained Realisations) which is used to generate initial conditions (ICs). To construct ICs we first use the Wiener filtered \citep[WF;][]{Zaroubi1995} to reconstruct the density field which is used to calculate the linear displacement field. The displacement field is used to push the constraint points to their locations in Lagrangian space at early times. This assumes linear perturbation theory and is referred to as the reverse Zeldovich approximation \citep[RZA;][]{Doumler1}. Constrained realisations are then generated \citep{CR1991, Doumler3} from the corrected location of galaxies after applying the RZA and the density scaled to an initial redshift to which ICs are generated. In this work we modify these methods to allow for MG ICs. We test and run constrained simulations using the COmoving Lagrangian Acceleration method \citep[COLA;][]{COLA}, specifically \texttt{MG-PICOLA} \citep{Winter2017} which implements the COLA method for a variety of MG models.

The paper is organised as follows, in Sec. \ref{theory} we discuss the theory behinds the models and constrained realisations, in Sec. \ref{method} we discuss the methodology for the current $\Lambda$CDM implementation and modification for MG, in Sec. \ref{results} we discuss the results on the constrained initial conditions and constrained COLA simulations and in Sec. \ref{conclusion} we discuss the results, future work and challenges.

\section{Theory\label{theory}}

\subsection{Cosmological models}

Here, we provide a brief description of the cosmological models we investigate. Our fiducial case that we take as a baseline with respect to which we measure all deviations and signals is the standard $\Lambda$CDM. We explore the physics of non-standard structure formation in two families of MG models, $f(R)$ and nDGP. For all the simulations we assume the same background cosmology based on \citep{Planck2020} with parameter values: $\Omega_{\rm m}=0.3111$, $\Omega_{\rm b}=0.049$, $\Omega_{\Lambda}=0.6889$, $H_{0}=67.66$,
$A_{\rm s}=2.105\times 10^{-9}$ and $n_{\rm s}=0.9665$ in a spatially flat universe (\ie{} $\Omega_{\rm k}=0$) and with neutrinos assumed to be massless (\ie{} $\sum m_{\rm \nu}=0\,{\rm eV}$). That is all models share the same expansion history, and the various cosmologies differ in the structure formation pace (\ie{} the linear growth rate) and the locally operating highly non-linear fifth-forces and screening mechanism (for the case of MG).

\subsubsection{Modified gravity models: nDGP and f(R)}

Observational constraints from the Solar System and massive bodies significantly reduce the possible extensions and deviations from Einstein's theory of General Relativity. Any modifications need to simultaneously match Solar System constraints for gravity but at the same time need to allow for possible departures from $\Lambda$CDM on large cosmological scales. Typically this is achieved in MG by the inclusion of a screening mechanism.

In nDGP, gravity is propagated through extra dimensions, unlike other forces, while in $f(R)$, non-linear functions are added to the Ricci scalar. In both cases, the new effective action integral will allow for extra degrees of freedom, that can be modelled by additional scalar fields and their interactions with matter. As a result, the new dynamics of these class of models allows for a non-vanishing fifth scalar-like force operating on cosmological and intergalactic scales. The physics of models from both families naturally contains a non-linear mechanism to suppress propagation of such fifth-forces. These are called screening mechanism, and in general the effective range of their operation is limited to small non-linear scales. In the nDGP model, screening is achieved by the means of the Vainshtein effect \citep{Vainshtein1972}; while in the $f(R)$ family, the corresponding mechanism is the Chameleon screening effect \citep{Chameleon2004}. The Vainshtein screening is dependent on the mass and distance from a screened object with no explicit dependence on either local or global environment, while the chameleon screening is dependent on the local curvature, and thus effectively on the density distribution of the local matter fields.

\begin{figure}
  \includegraphics[width=\columnwidth]{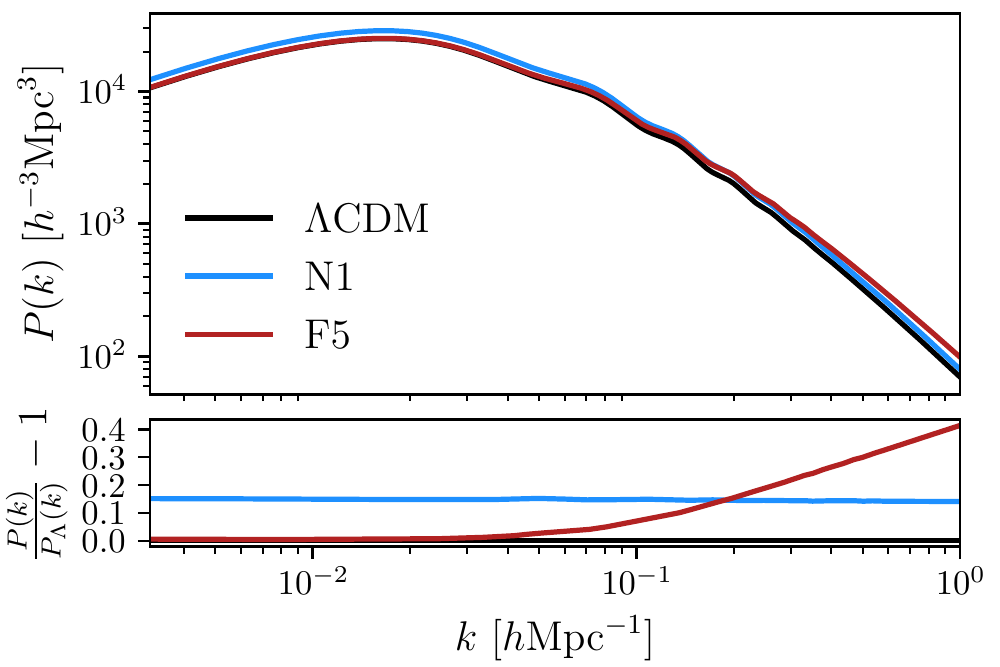}
  \caption{The linear power spectra at redshift $z=0$ is shown for the fiducial $\Lambda$CDM model, the nDGP model N1, and the $f(R)$ model F5 in the top panel. Differences with respect to $\Lambda$CDM are shown in the bottom panel. \label{fig_pktheory}}
\end{figure}

As a test-bed for the MG theories we select one particular model for nDGP and one model for $f(R)$. In both cases we assume a background cosmology following the parameters given for the fiducial $\Lambda$CDM model discussed above with the additional parameter $r_{\rm c}H_{0}=1$ for nDGP and $|f_{R0}|=10^{-5}$ for $f(R)$. We refer to these models as N1 and F5, respectively. At high redshift these models exhibit almost identical clustering properties as $\Lambda$CDM but depart at low redshift. This is showcased by the differences in their linear power spectra at redshift $z=0$ (shown in Fig.~\ref{fig_pktheory} and obtained from a modified version of CAMB) where nDGP shows a scale independent shift, while in $f(R)$ the shift is scale dependent -- with large scales consistent with $\Lambda$CDM but departing at smaller scales.

\subsection{Constrained simulations}
\label{subsec:CRS}

The motivations for constrained simulations are many. One particular goal is to create simulations that mimic properties of data and the view of our local Universe that it offers. Another important aspect of constrained simulations is that they allow for studying, and thus limiting to some extent, the impact of cosmic variance on nearby non-linear structures. In this context, it is crucial to remember that the observational galaxy data is noisy and sparse. Thus the constrained simulations both replicate properties of the data but also add information where data is missing. 

Our approach involves three major steps: (1) WF which reconstructs the linear velocity and density field from peculiar velocity measurements; (2) the RZA which reduces linear-order shifts in the position of constraints relative to an initial redshift, and (3) the generation of constrained realisations which adds random fluctuations in places which are poorly constrained. It is important to note that constrained realisations assume the input data and realisations are Gaussian. Of course low redshift observations of large scale structure are certainly non-Gaussian, so some care is needed to try to limit measurements to the linear regime. The latter is usually attained by considering all the relevant density and velocity fields smoothed on scales where the effects of the non-linearities are already reduced. The procedure for obtaining constrained field realisations using the observed velocity data is described in detail in \citep{Zaroubi1999}. Below, for completeness, we discuss the main steps of this method.

\begin{figure*}
  \includegraphics[width=\textwidth]{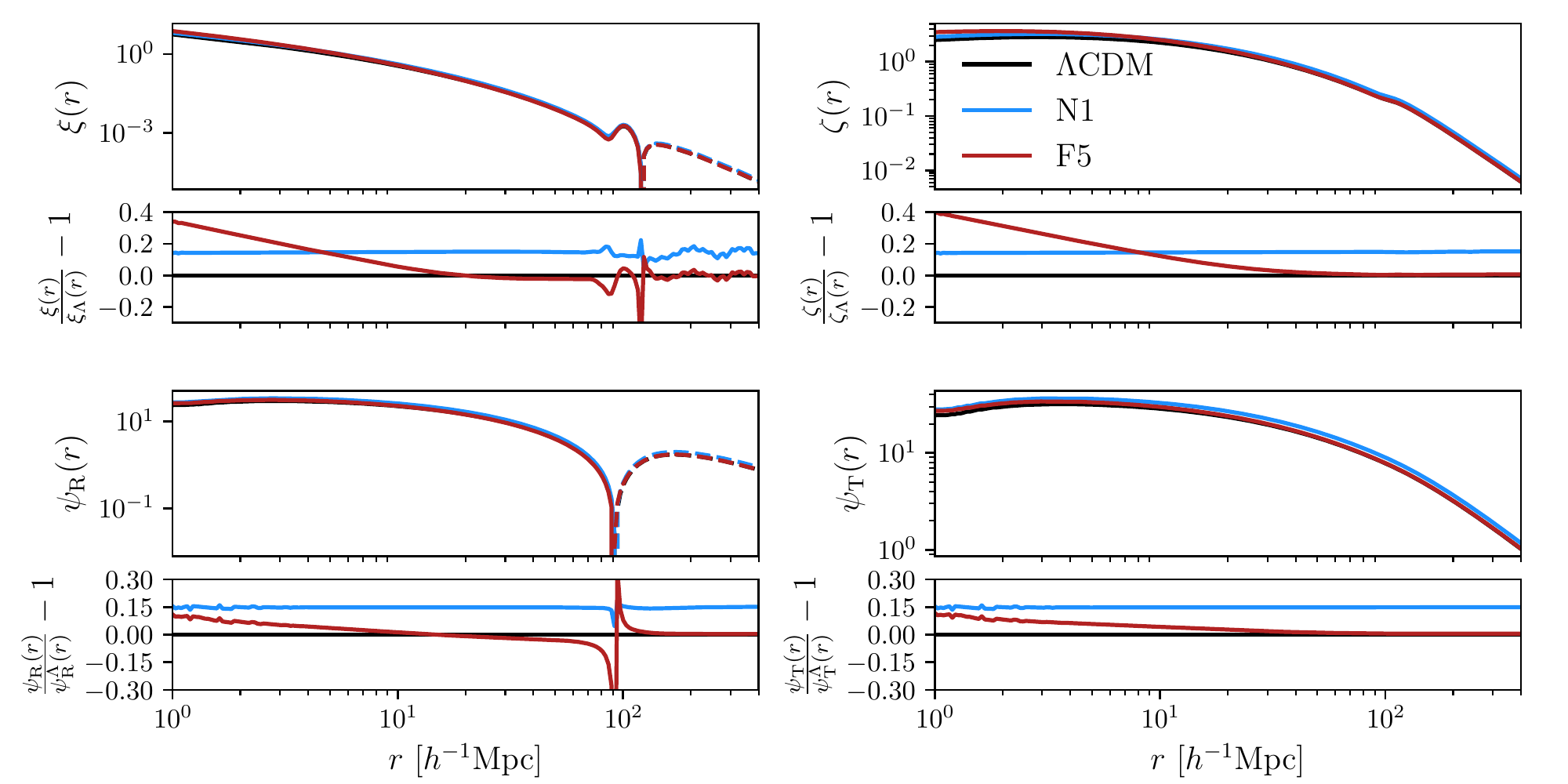}
  \caption{The density and velocity correlation functions are shown for the fiducial $\Lambda$CDM model and the MG models of nDGP N1 and $f(R)$ F5. In the top left panel $\xi$ is the density-density correlation function, in the top right panel $\zeta$ is the velocity-density cross correlation function and in the bottom panels are the velocity-velocity correlation functions ($\psi_{\rm R}$ the radial component on the left and $\psi_{\rm T}$ the tangential component on the right). Subpanels show the differences between these correlation functions with respect to the base fiducial $\Lambda$CDM model.\label{fig_correlators}}
\end{figure*}

\subsubsection{Wiener filtering}

Following \citep{Zaroubi1999} the WF density $\delta^{\rm WF}$ is estimated from a given set of peculiar velocity $\pmb{u}$ constraints from the relation
\begin{equation}
  \delta^{\rm WF}(\pmb{r}) = \sum_{i=1}^{M}\sum_{j=1}^{M}\langle \delta(\pmb{r})\,\pmb{u}_{i}\rangle\, \langle \pmb{u}_{i}\,\pmb{u}_{j}\rangle^{-1}\,\pmb{u}_{j}\,,
\end{equation}
where $\langle \delta(\pmb{r})\,\pmb{u}_{i}\rangle$ is given by the linear theory velocity-density correlation, $\langle \pmb{u}_{i}\,\pmb{u}_{j}\rangle$ is the peculiar velocity covariance matrix and the indices $i$ and $j$ sum over $M$ peculiar velocity constraints. The density-velocity correlation vector is defined as
\begin{equation}
  \label{eq_cross}
  \langle \delta(\pmb{r})\,\pmb{u}_{i}\rangle = -\dot{a}f\zeta(r')\left(\pmb{e}_{i} \cdot \pmb{\bar{r}}'\right),
\end{equation}
where $a$ is the scale factor related to the redshift $z$ by $a=1/(1+z)$, $\dot{a}=a\,H(z)$ and $H(z)$ is the Hubble parameter (note at $z=0$ this reduces to $\dot{a}=H_{0}$). The logarithmic growth rate $f={\rm d}\ln D/ {\rm d}\ln a$ where $D$ is the linear growth function normalised at $z=0$ and $\zeta(r)$ is the density-velocity correlation function. Lastly $r'=|\pmb{r}'|$, $\pmb{\bar{r}}'=\pmb{r}'/r'$ and $\pmb{e}_{i}$ is a unit vector describing the direction of the peculiar velocity constraint $\pmb{u}_{i}$, in this study this is equivalent to the line-of-sight vector.

The velocity covariance matrix is given by
\begin{equation}
  \langle \pmb{u}_{i}\,\pmb{u}_{j}\rangle = \pmb{e}_{i}\cdot\langle \pmb{u}_{i}(\pmb{r})\,\pmb{u}_{j}(\pmb{r}+\pmb{r}')\rangle\cdot\pmb{e}_{j} + \sigma_{i}^{2}\delta^{\rm K}_{ij},
\end{equation}
where $\sigma_{i}$ are measurement errors on the peculiar velocities, $\delta^{\rm K}_{ij}$ is the Kronecker delta function and the the components of the linear theory peculiar velocity covariance matrix (\ie{} velocity correlation tensor) are given by
\begin{equation}
  \label{eq_cov}
  \begin{split}
    \Psi^{v}_{\alpha\beta}\equiv\langle \pmb{u}(\pmb{r})\, & \pmb{u}(\pmb{r}+\pmb{r}')\rangle_{\alpha\beta}\, = (\dot{a}f)^{2}\\
    &\times\bigg\{\psi_{\perp}(r')\delta^{K}_{\alpha\beta} + \big[\psi_{\parallel}(r') - \psi_{\perp}(r')\big]\bar{r}'_{\alpha}\bar{r}'_{\beta}\bigg\}.
  \end{split}
\end{equation}
Where $\psi_{\perp}(r)$ and $\psi_{\parallel}(r)$ are the tangential and radial velocity-velocity correlation functions. The above decomposition of the full velocity correlation into its radial and tangential holds for a statistically homogeneous and isotropic velocity field \citep{Gorski1988}. We can define the auto- and cross-correlation functions for the density and peculiar velocities as follows
\begin{align}
  &\xi(r) = \frac{1}{2\pi^{2}}\int_{0}^{\infty} k^{2}\, P(k)\, j_{0}(kr)\, {\rm d}k\,,\label{eq_xi}\\
  &\zeta(r) = \frac{1}{2\pi^{2}}\int_{0}^{\infty} k\, P(k)\, j_{1}(kr)\, {\rm d}k\,,\label{eq_zeta}\\
  &\psi_{\parallel}(r) = \frac{1}{2\pi^{2}}\int_{0}^{\infty}\left(P(k)\, j_{0}(kr) -2\, \frac{j_{1}(kr)}{kr}\right)\, {\rm d}k\,,\label{eq_psiR}\\
  &\psi_{\perp}(r) = \frac{1}{2\pi^{2}}\int_{0}^{\infty}P(k)\, \frac{j_{1}(kr)}{kr}\, {\rm d}k\,\label{eq_psiT},
\end{align}
where $\xi(r)$ is the density-density correlation function. The integrals are functions of the linear matter power spectra $P(k)$ and spherical Bessel functions $j_{n}(x)$ (\ie{} a spherical Bessel function at $x$ of order $n$). In the above relation, we will be using consistently $P(k)$ and $f$ of either $\Lambda$CDM or a given MG model respectively. The numerically computed correlation functions are shown in Fig.~\ref{fig_correlators} for two MG models and $\Lambda$CDM. Negative correlations are seen for large scales in $\xi$ and $\psi_{\rm R}$ and the wiggles seen in the subpanel comparisons of $\xi$ originate from numerical instabilities at larger radii where the correlation values are small.

\subsubsection{Reverse Zeldovich approximation}

If constrained simulations were generated directly from the peculiar velocity constraints at their native (\ie{} final observed) positions, the simulations generated would have structures that are systematically shifted from the input field. This is because the location of structures move with respect to an arbitrary point of origin as they grow and evolve. This is a direct consequence of large and small-scale coherent (\ie{} bulk) flow motions in the Universe. In order to account for this and to ensure that the final simulated structures appear at the comoving positions corresponding with the observational data locations we need to push our constraints to where they would appear at early times \ie{} the initial redshift. In \citep{Doumler1} the dominant effect was shown to be well captured already by the terms from linear perturbation theory. Thus it is sufficient to perform a simple linear order shift to the positions of the peculiar velocity constraints, known as the Reverse Zeldovich approximation (RZA) \citep{Nusser1994}.

The Zeldovich approximation \citep{ZA1970}, based on linear-order perturbation theory, can be used to describe the relation between the initial (\ie{} Lagrangian) position of a fluid element in space, with its final (\ie{} Eulerian) position as a function of time
\begin{equation}
  \pmb{x}(\pmb{q},z) = \pmb{q}  + \pmb{\psi}(\pmb{q}, z)\,,
\end{equation}
where $\pmb{x}$ is the final position in space, $\pmb{q}$ the initial condition position and $\pmb{\psi}$ the displacement field, all given in comoving coordinates. The peculiar velocities $\pmb{u}$ are related to the displacement field by
\begin{equation}
  \label{eq_vel_vs_dis}
  \pmb{u}(\pmb{q}, z) = \dot{a}\, f\,\pmb{\psi}(\pmb{q}, z)\,.
\end{equation}
In the RZA the following approximation is employed,
\begin{equation}
  \pmb{q}_{\rm RZA}(\pmb{x}) \sim \pmb{x}(z) - {\pmb{v}(\pmb{x}, z)\over\dot{a} f}\,,%{\dot{a}\, f}\,,
\end{equation}
to give the initial condition positions of the peculiar velocity constraints. Note, in our work $\pmb{\psi}$ is the linear displacement computed from the reconstructed WF velocity field.

\subsubsection{Constrained Realisations}

The final step is to now construct specific initial condition realisations from the constraints (\ie{} the data) after performing the RZA. This creates realisations which retain information at locations with high constraints, but with added noise and random information at scales and locations which are poorly constrained. The density contrast for a constrained realisation (CR) is obtained from the following relation
\begin{equation}
  \begin{split}
  \delta^{\rm CR}(\pmb{r}) &= \delta^{\rm RR}(\pmb{r}) \\
    &+ \sum_{i=1}^{M}\sum_{j=1}^{M}\langle \delta(\pmb{r})\,\pmb{u}_{i}\rangle\, \langle \pmb{u}_{i}\,\pmb{u}_{j}\rangle^{-1}\,(\pmb{u}_{j} - \pmb{u}^{\rm RR}_{j})\,,
  \end{split}
\end{equation}
following \citep{CR1991} where $\pmb{u}$ are the peculiar velocities at their initial condition positions (i.e. after applying the RZA),  $\delta^{\rm RR}$ the density from a random realisation and $\pmb{u}^{\rm RR}$ the peculiar velocity from the random realisation at the position and direction of the peculiar velocity constraints. The density and displacement field are related in Fourier space by
\begin{equation}
  \pmb{\psi}(\pmb{k}) = \frac{i\pmb{k}}{k^{2}}\delta(\pmb{k}),
\end{equation}
since $\nabla^{2}\Phi=\delta$ and $\pmb{\psi}=-\pmb{\nabla}\Phi$, and where $k=|\pmb{k}|$ and $i$ is the imaginary unit.

\subsubsection{\texttt{ICeCoRe} $f(R)$ implementation}

The production of constrained realisations from \texttt{ICeCoRe} broadly follows the steps outlined above. However, \texttt{ICeCoRe} was designed for constructing constrained realisations in $\Lambda$CDM and as a result there are a few design choices which are $\Lambda$CDM-specific. Although these limitations do not limit the accuracy of constrained nDGP simulations they do affect the accuracy of constrained $f(R)$ simulations. The most important of them is the choice to take constraints of the displacement field rather than the velocity field directly. The two are linked by  Eq.~\eqref{eq_vel_vs_dis} where a simple division by the factor $\dot{a}f$ can convert $\pmb{u}$ to $\pmb{\psi}$, but this is only true when $f$ is scale independent. To correctly convert one to the other in $f(R)$ would require a multiplication in Fourier space, but this is impractical given the constraints $\pmb{u}$ do not lie on a uniform grid (where we can take advantage of Fast Fourier Transforms -- FFT and numerical methods). The choice to constrain the displacement field also leads to the simplification of Eq.~\eqref{eq_cross} and Eq.~\eqref{eq_cov} where the $\dot{a}f$ terms drop from the equation. To generate self-consistent $f(R)$ constrained simulations will require developing \texttt{ICeCoRe} with a different philosophy. This will require moving the linear growth function $f$ from Eqs.~\ref{eq_cross} and \ref{eq_cov} to inside the integrals of Eqs.~(\ref{eq_zeta}--\ref{eq_psiT}). Furthermore, we will need to use the peculiar velocity constraints directly to generate the constrained realisations rather than converting to the displacement field. Implementing this change will be the subject of future work. For the moment we instead consider an effective scalar value for $f$ at some chosen scale.

\section{Method\label{method}}

In this section we discuss the steps for constructing constrained simulations for $\Lambda$CDM and MG. We begin by discussing the peculiar velocity constraints used, then discuss the implementation of \texttt{ICeCoRe} for generating initial conditions for $\Lambda$CDM followed by a discussion of the modifications to the \texttt{ICeCoRe} inputs and intermediate procedures for generating initial conditions for MG. Lastly, we discuss the subsequent COLA $N$-body simulations and the random realisations used for comparisons.

\subsection{Peculiar velocity constraints from CosmicFlows-2}

We use CosmicFlows-2 (CF2) \citep{CF2}, a catalog of galaxy redshifts and distances, and hence of galaxy distances and peculiar velocities, for imposing the constraints. The data consist of measurements of the sky positions, distance moduli and redshifts for over 10,000 local Universe galaxies and groups of galaxies, which are mostly contained within a distance of $\lesssim 150\, h^{-1}{\rm Mpc}$. Galaxies which are closely clustered are grouped and the peculiar velocity constraints replaced by their mean -- suppressing virial motions in high density environments \citep{Sorce2017} and maintaining the suitability of linear theory assumptions. 
%\yh{({\it YH: Note distances are calculated assuming $H_0=75\,km/s/Mpc$ })
A straight forward transformation from the noisy distance muduli and redshifts to distances and velocities induces Malmquist-like biases. The data used here is corrected by the bias minimization algorithm \citep{Sorce2015}.   

Although newer CosmicFlows data releases are available \citep{CF3, CF4}, CF2 \citep{CF2} has been widely used in the community for the production of constrained simulations by CLUES \citep{Sorce2016} and systematics are well understood \citep{Sorce2015}, while the effects in later releases still require further study. As we are simply interested in comparing and studying the implementation of MG for constrained simulations the most up-to-date data are not required for the time being.

\subsection{Constrained realisations for $\Lambda$CDM}
%\WH{WOJMARK -don't delete}
Constrained simulations are generated by \texttt{ICeCoRe} in four steps. In each step \texttt{ICeCoRe} takes as input the linear power spectrum $P(k)$ which is computed for $\Lambda$CDM using the cosmological Boltzmann code solver \texttt{CAMB} \footnote{\url{https://camb.readthedocs.io/}}.
\begin{enumerate}
  \item Wiener Filtering: Peculiar velocity constraints from CF2 are used to construct the WF density field and velocity field computed by differentiating the WF density field in Fourier space.
  \item Reverse Zeldovich Approximation: Using interpolated values of the displacement field, from the WF reconstruction, we subtract the linear-order displacement to place the constraints to their positions in the initial conditions in Lagrangian space.
  \item Constrained realisations: A random seed is given to generate a CR of the density field using the peculiar velocities at the RZA positions following the density power spectrum at $z=0$.
  \item Initial conditions: The density field at the initial conditions are calculated by multiplying the CR density by the linear growth function at the initial redshift $z_{\rm IC}=49$. IC particles are then generated on a grid using the Zeldovich approximation.
\end{enumerate}

\begin{figure}
  \includegraphics[width=\columnwidth]{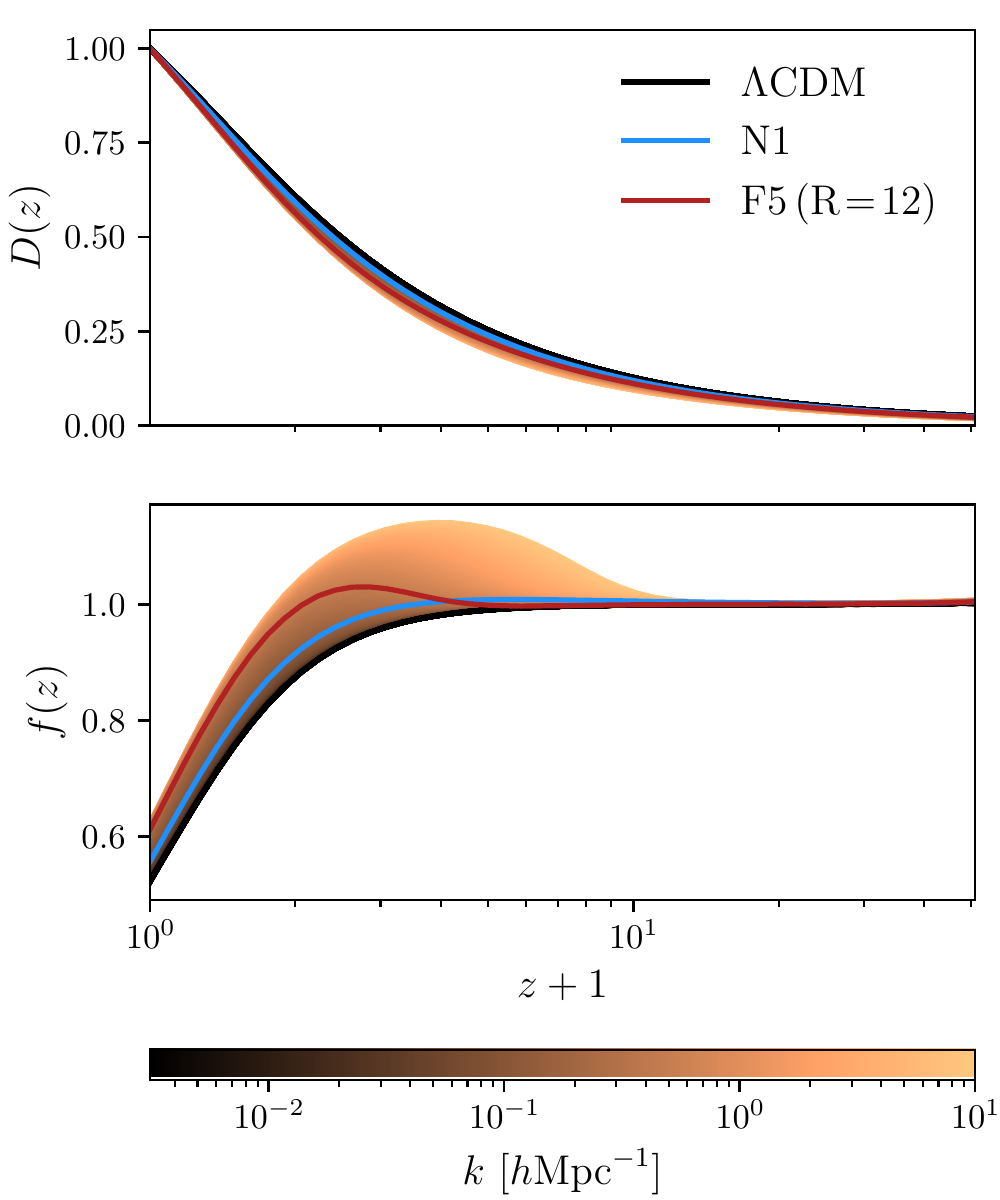}
  \caption{The scale-independent growth functions $D(z)$ and $f(z)$ are shown for the fiducial base $\Lambda$CDM model (in black), the nDGP model N1 (in blue) and the scale-dependent growth functions shown for the $f(R)$ model F5 (where colours correspond to scales $k$ in the colourbar) and the red line shows the effective growth functions at a scale of $k=2\pi/12\,{h{\rm Mpc}^{-1}}$.\label{fig_num_growth}}
\end{figure}

\subsection{Constrained realisations for modified gravity}

Constrained simulations for MG are implemented with \texttt{ICeCoRe} in a similar method to $\Lambda$CDM. We take advantage of the modular set up to alter the inputs and intermediate steps for MG and compute scale independent (for nDGP) and dependent (for $f(R)$) growth functions numerically from MG power spectra.

The first modification is the input of a MG power spectra. These are computed using \texttt{MGCAMB} \footnote{Made available to us by Hans Winter} a modified version of \texttt{CAMB} used to compute power spectra for a range of MG models. The second step requires the pre-calculation of growth functions. Rather than hard-code this procedure into our pipelines, we compute growth functions completely numerically from linear power spectra evaluated at many redshifts. The linear scale-dependent growth function is calculated by taking
\begin{equation}
  D(k, z) = \sqrt{\frac{P(k, 0)}{P(k, z)}},
\end{equation}
and the logarithmic growth function computed by numerical differentiation,
\begin{equation}
  f(k, z) = \frac{{\rm d}\log D(k, z)}{{\rm d}\log a}.
\end{equation}
Note, for the above we first interpolate $D(k, z)$ onto a regular $\log a$ grid using a cubic spline, since numerical differentiation performs best on a regular grid. For the scale-independent case we take the mean of $k$ above superhorizon scales, \ie{} $0.01<k<10\, h{\rm Mpc}^{-1}$ (since we are using the linear power spectra the inclusion of non-linear scales has no effect on the linear growth functions except to provide more stable and accurate results). In Fig.~\ref{fig_num_growth} we show the numerically calculated growth functions for nDGP (scale-independent) and $f(R)$ (scale-dependent).

\begin{figure*}
	\includegraphics[width=\textwidth]{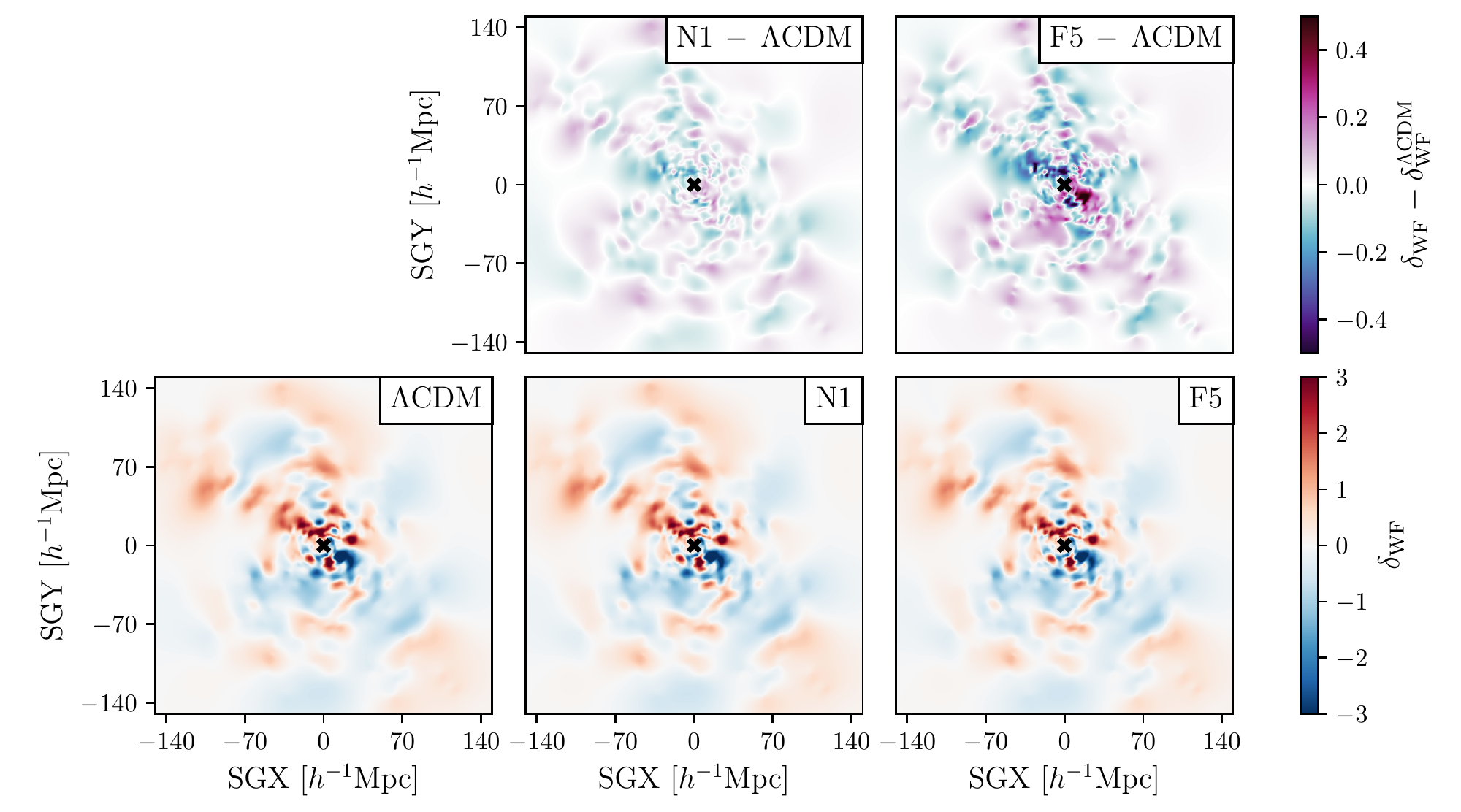}
	\caption{The Wiener filtered (WF) reconstruction of the density field from peculiar velocity data from CosmicFlows-2 is shown for $\Lambda$CDM on the bottom left, the N1 nDGP model on the bottom middle and the F5 $f(R)$ model on the bottom right panel. In the top panels we show deviations of the WF reconstruction of N1 (middle) and F5 (right) from $\Lambda$CDM. The WF recovers the linear density field extrapolated to the present epoch, hence the unphysical values below $-1$. The maps are shown with respect to the supergalactic $x$ (SGX) and $y$ (SGY) coordinate axis for a slice of width $20\,{h^{-1}{\rm Mpc}}$ in the supergalactic $z$ axis (\ie{} $-10\leq {\rm SGZ} \leq 10 \,{h^{-1}{\rm Mpc}}$). The WF reconstruction show fairly consistent structures with the strongest deviations seen from F5.\label{fig_WF}}
\end{figure*}

The steps to produce constrained realisations for MG follow the procedure outlined for $\Lambda$CDM with some important differences: such as the use of MG power spectra and numerically computed growth functions. Note, for $f(R)$ we input growth functions at a single scale, $k=2\pi/12\, h{\rm Mpc}^{-1}$, since scale dependent growth functions are currently not implemented in \texttt{ICeCoRe}. This choice of scale is motivated by the effective smoothing of the Wiener filter reconstruction (of the density field) from peculiar velocities which is roughly on the scale of $12\,h^{-1}{\rm Mpc}$. The other additional change we make is to rescale the CR densities to the IC redshift using scale-dependent growth functions via FFT, which is implemented outside of \texttt{ICeCoRe}.

\begin{figure*}
    \includegraphics[width=\textwidth]{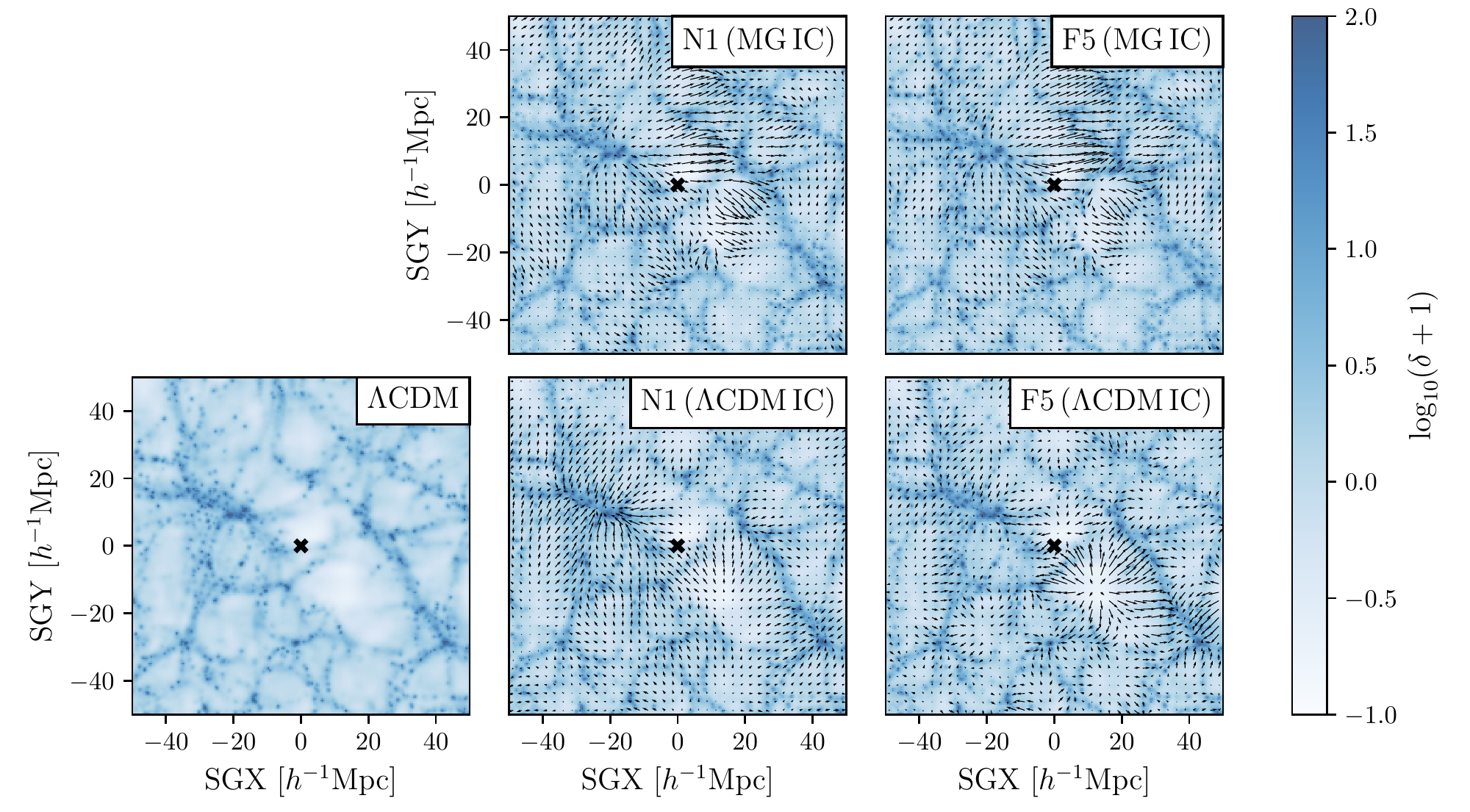}
    \caption{The density of constrained simulations for $\Lambda$CDM and MG are shown in a region of 100 $h^{-1}{\rm Mpc}$ around the observer. On the bottom panels are constrained simulations all generated with $\Lambda$CDM ICs while on the top are generated with MG ICs. The difference between the structures from different models is hard to distinguish by eye; for visual comparisons we indicate the differences between positions of matched particles with respect to $\Lambda$CDM with arrows/quivers.}
    \label{fig:dens_comparison}
\end{figure*}

\subsection{Constrained and random simulations}

To test the effect of applying constraints for different models we also construct initial conditions with the same random seeds with the corresponding $\Lambda$CDM or MG power spectra. Once these are produced we then use the initial condition particles to run COLA simulations using \texttt{MG-PICOLA} which implements the COLA method for the MG models of nDGP and $f(R)$ \citep{Winter2017}. To be able to construct halo catalogues from COLA requires that the particle-mesh grid be a factor of $>3$ times the number of particles along a single axis. For most of the simulations our initial conditions are generated with $512^{3}$ particles over a $500\, h^{-1}{\rm Mpc}$ box, so for added validity and due to the relative inexpensive computation cost of running COLA we use a particle-mesh grid size of $2048^{3}$. For each model we produce 5 constrained initial conditions and 5 random initial conditions with the same random seeds. Constrained simulations are then computed with \texttt{MG-PICOLA} for each model. For MG models these are run from both $\Lambda$CDM and MG initial conditions. In total this means the construction of 30 initial conditions from which 50 simulations were produced.

\section{Results\label{results}}

In this section we compare results obtained from constrained initial conditions from $\Lambda$CDM and MG models nDGP and $f(R)$; including the WF reconstruction and COLA simulation outputs and statistics.

\subsection{Comparing Wiener filtered reconstruction}

The WF reconstruction of the density field computed by \texttt{ICeCoRe} is shown in Fig.~\ref{fig_WF}. We see the fields are strikingly similar, with the differences originating from  differences in the model's power spectra and correlation functions (shown in Fig.~\ref{fig_pktheory} and \ref{fig_correlators}, respectively). For nDGP the variations appear equal in amplitude throughout the volume, perhaps unsurprising given the differences between the power spectra of $\Lambda$CDM can be described by an effective amplitude shift on all scales. For $f(R)$ the differences to $\Lambda$CDM are strongest closer to the observer, showing a strong relation to the amplitude of the density field.\\
%\it{YH: I'm not sure about your interpretation. The variation of the amplitude of the F5-$\Lambda$CDM does not depend on the scale - in the sense of Fourier decomposition - but on the proximity to the observer. Nearby structures of the residual field have higher amplitudes. My reading of it is that the amplitude of the residual depends on the amplitude of the target field.}

\subsection{Comparing COLA simulation statistics}

\begin{figure*}
    \includegraphics[width=\textwidth]{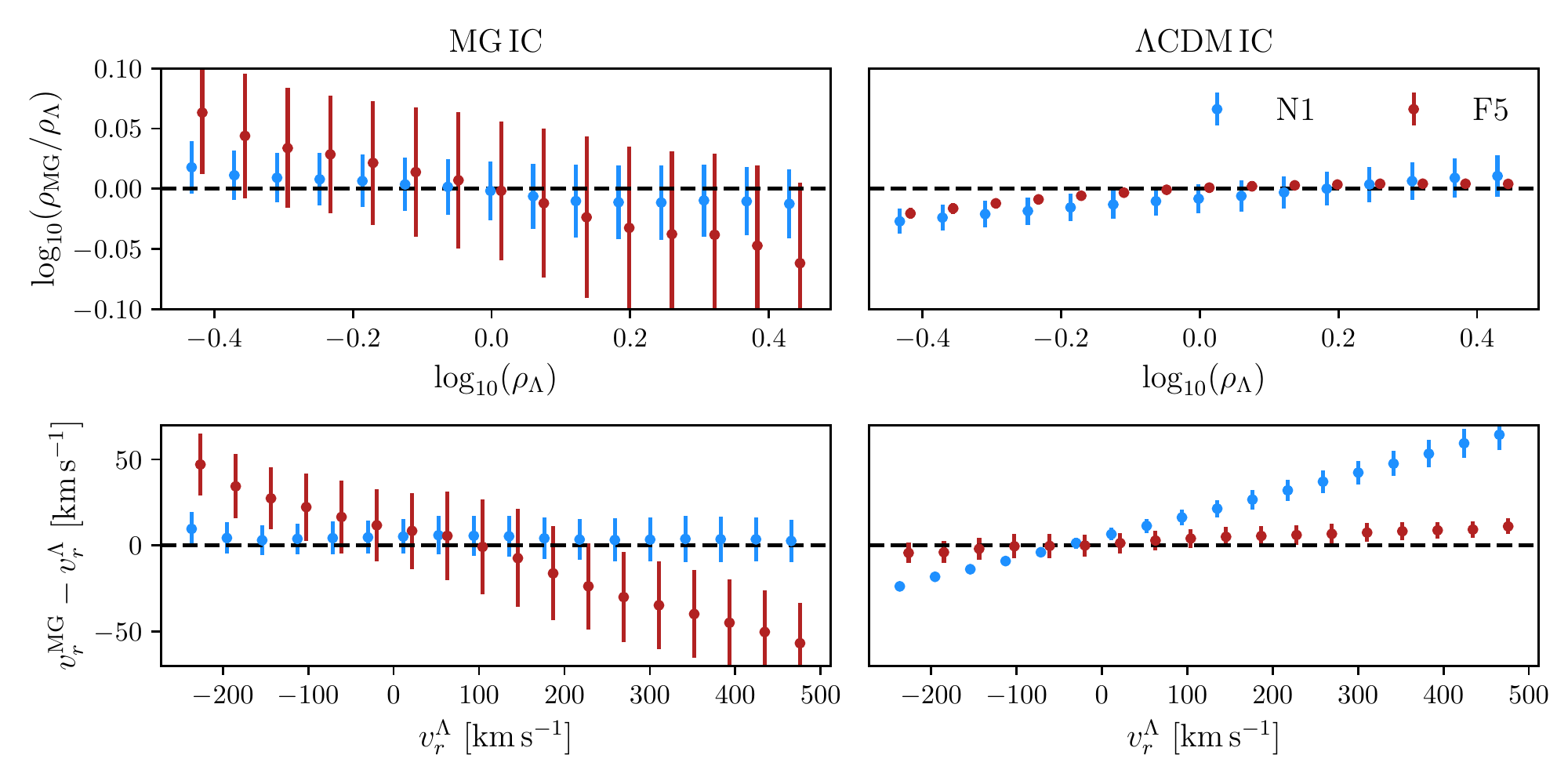}
    \caption{The density (top) and radial velocity (bottom) fields are compared between MG and $\Lambda$CDM constrained simulations with MG ICs (left) and $\Lambda$CDM ICs (right). The density and velocity fields are computed on a $256^{3}$ grid using the triangular-shaped-cloud mass-assignment scheme and compared within a radius of $30\,h^{-1}{\rm Mpc}$ from the origin \ie{} where most of the CF2 constraints are located. On the y-axis we show the difference in densities between $\Lambda$CDM and MG on the top row and radial velocities on the bottom row.}
    \label{fig:dens_vel_comp}
\end{figure*}

To test the validity of the constrained simulation methodology for MG we began by comparing the power spectra and halo mass functions computed from the simulations. These were found to be completely consistent between constrained and random simulations. Since these statistics are insensitive to phases, this result is not surprising and we move onto comparisons of the density and velocity fields.

In Fig.~\ref{fig:dens_comparison} we compare the constrained simulation density fields for the first realisation. Particles within $\pm10\,h^{-1}{\rm Mpc}$ from ${\rm SGZ}=0$ at redshift $z=0$ are used to construct a 2-dimensional projection of the density on a slice of width $100\,h^{-1}{\rm Mpc}$ around the origin. The simulations show that the different models and choice of initial conditions have small yet subtle effects, with structures between simulations seeming to be rather similar. To aid the eye we indicate the differences between matched particles from $\Lambda$CDM to MG simulations with arrows/quivers. For the $\Lambda$CDM ICs the MG simulations show more extreme features, \ie{} denser clusters and more emptier voids an effect driven by the larger clustering properties for these models. However, for the MG ICs the differences are not as trivial, with small displacements near clusters but large displacements between structures. We interpret this as the constraints dictating the final location of the main structures; leading to differences in the field between structures owing to the model's different formation histories.

To get a better comparison we compute the density and velocity field on a $256^{3}$ grid using the triangular-shaped-cloud particle mass-assignment scheme. We then isolate the cells within a $30\,h^{-1}{\rm Mpc}$ radius \citep[similar to][]{Sorce2014} and compare the densities and radial velocities cell-by-cell shown in Fig~\ref{fig:dens_vel_comp}. Constrained simulation in $\Lambda$CDM, following the procedure in this paper, have previously been shown to be consistent with input peculiar velocities \citep[see][]{Sorce2014}. Therefore, by comparing to the $\Lambda$CDM peculiar velocities and density fields we can test the accuracy of the MG constrained simulations.

For MG constraints simulations produced with $\Lambda$CDM ICs we find the density and radial velocities to be systematically shifted with low variance. For the nDGP simulations produced with MG ICs we find consistent radial velocities with $\Lambda$CDM. For $f(R)$, with MG ICs, we see a strong negative correlation between radial velocities. Since the input are peculiar velocities we expect these values to be consistent between models. This shows the MG ICs for nDGP are performing as we expect and this method can be used for future nDGP constrained simulations produced at higher resolution. However, for $f(R)$ we find large discrepancies between the simulations produced with MG ICs and find more consistent radial velocities with $\Lambda$CDM ICs; although the latter comes at the cost of a systematic shift. Since on large scales $f(R)$ has consistent growth functions as $\Lambda$CDM this results shows that, in hindsight, our choice of scale was slightly misguided and that we ought to have chosen a value at a larger scale, which more closely reflect the scales that are more influential in structure formation. For the time being, constrained simulations for $f(R)$ are better produced with $\Lambda$CDM ICs rather than the approximate methods currently used for MG ICs.

\section{Conclusions\label{conclusion}}

%\WH{I think it is paramount to add in the conclusions a short paragraph  about our Wiener reconstruction results. First time-ever Wiener done for non-LCDM cosmology.}
In this paper, we extend the methodology for constructing constrained simulations of the local Universe in $\Lambda$CDM to the MG models nDGP and $f(R)$. We begin by describing the formal extensions to the methodology; the input of different power spectra, correlation functions, and the numerical calculation of growth functions. We use the \texttt{ICeCoRe} package \citep{Doumler3}, previously used by CLUES \citep{CLUES2010} and HESTIA \citep{HESTIA2020} for constructing the initial conditions for constrained simulations in $\Lambda$CDM. Our implementation makes use of the modular procedures to incorporate steps for MG. However, while this implementation is completely compatible with the scale-independent growth functions of nDGP, the implementation for $f(R)$ is incomplete. This is due to the scale-dependence of the growth functions which are currently not implemented in \texttt{ICeCoRe}.
This issue arises due to a design choice in \texttt{ICeCoRe} to constrain the density field from the displacement field and not the peculiar velocities directly. The conversion from velocity to displacement includes a division by the logarithmic growth function; for the scale-independent case this is a simple conversion but for $f(R)$ this multiplication needs to be carried out in Fourier space. For the peculiar velocity constraints applying this conversion is impractical and since conducting this in a fully self-consistent way would require rewriting significant portions of \texttt{ICeCoRe} we have opted for an approximate scheme: taking an effective $f$ at a scale of $k=2\pi/12$ $h{\rm Mpc}^{-1}$ (a real space scale of $R=12$ $h^{-1}{\rm Mpc}$) roughly corresponding to the smoothing incurred from reconstructing the density from peculiar velocities assuming linear theory.

We construct WF reconstructions of the density field from the CF2 peculiar velocities for $\Lambda$CDM and MG nDGP and $f(R)$ models -- the first such construction for non-$\Lambda$CDM models. The WF reconstruction, on the whole, show very similar features with subdominant variations owing to differences in the MG linear power spectra; in nDGP features are more prominent at all scales while for $f(R)$ this is limited to the largest amplitude features closest to the origin (\ie{} Local Group). 

ICs for five constrained realisations were then generated with corresponding ICs for random realisations with matching seeds. The WF reconstruction of the mean density field show very similar reconstructions with subdominant variations owing to differences in the MG linear power spectra. COLA MG simulations were generated using both the $\Lambda$CDM and MG ICs.

Comparisons of the density field show that using $\Lambda$CDM ICs will generate structures for MG that are more clustered and voids that are emptier, this is incompatible with the constraints from peculiar velocity data and shows the necessity for generating self-consistent MG ICs. Here we show that the nDGP MG ICs reproduce consistent radial velocities with $\Lambda$CDM, a property we expect since the constraints are from peculiar velocities. For $f(R)$ the MG ICs are not fully self-consistent due to the scale-dependence of the growth functions and the approximations used. As a result the $f(R)$ simulations with MG ICs produce radial velocities that are inconsistent with $\Lambda$CDM. For $f(R)$ we find greater consistency with $\Lambda$CDM ICs. A result showing that our choice of scale for the approximation of the growth functions should have been larger, to reflect the scales which are more important for structure formation. For the time being this means that constrained simulations for $f(R)$ are more consistently produced with $\Lambda$CDM ICs until a fully self-consistent scale-dependent methods can be constructed.

In this paper we have extended the methodology for constructing constrained simulations to MG. We have shown the importance of conducting this in a fully self-consistent way (rather than using $\Lambda$CDM ICs), since MG imply subtle but important differences in the reconstructed WF field and imply different growth histories. Future work will look to extend this formalism to fully incorporate the scale-dependence of models such as $f(R)$ without the need to assume effective values for the growth rate. This will facilitate the future production and study of high-resolution and hydrodynamic MG simulations enabling the study of MG on properties of the local Universe on small scales.

% If you have acknowledgments, this puts in the proper section head.
\begin{acknowledgments}
We thank Adi Nusser for providing useful comments and discussions. The research conducted for this work was supported by the Polish National Science Centre grants no: 2018/31/G/ST9/03388, 2018/30/E/ST9/00698, 2020/39/B/ST9/03494, 2020/38/E/ST9/00395, and by the Polish Ministry of Science and Higher Education through grant DIR/WK/2018/12. NIL and SP acknowledges support from the Deutsche Forschungs Gemeinschaft joint Polish-German research project LI 2015/7-1. YH has been partially supported by the Israel Science Foundation grant ISF 1358/18.
\end{acknowledgments}

% Create the reference section using BibTeX:
\bibliography{bibfile}

\end{document}